\documentclass[%
 reprint,
 amsmath,amssymb,
 aps,
]{revtex4-1}

\usepackage{graphicx}
\usepackage{dcolumn}
\usepackage{bm}
\usepackage{enumerate}
\usepackage{color}

\begin{document}

\title{Efficient Low-Order Approximation of First-Passage Time Distributions}

\author{David Schnoerr$^1$}
\author{Botond Cseke$^3$}
\author{Ramon Grima$^2$}
\author{Guido Sanguinetti$^{1,}$}\email{gsanguin@inf.ed.ac.uk}
\affiliation{%
 {\mbox{$^1$School of Informatics, University of Edinburgh, Edinburgh, UK} \\
 \mbox{$^2$School of Biological Sciences, University of Edinburgh, Edinburgh, UK}}\\
 \mbox{$^3$Microsoft Research, Cambridge, UK}
}%

\begin{abstract}
We consider the problem of computing first-passage time distributions for reaction processes modelled by master equations. We show that this generally intractable class of problems is equivalent to a sequential Bayesian inference problem for an auxiliary observation process.  The solution can be approximated efficiently by solving a closed set of coupled ordinary differential equations (for the low-order moments of the process) whose size scales with the number of species. We apply it to an epidemic model and a trimerisation process, and show good agreement with stochastic simulations.
\end{abstract}

\pacs{Valid PACS appear here}

\maketitle

Many systems in nature consist of stochastically interacting agents or particles. Such systems are {frequently} modelled as {reaction processes {whose dynamics are described} by master equations \cite{Gardiner2009}}.  There are several examples of stochastic modelling { of reaction processes} in the fields of systems biology \cite{eldar2010functional, grima2009noise}, ecology \cite{mckane2005predator}, epidemiology \cite{dykman2008disease}, social sciences \cite{fernandez2014voter} and neuroscience \cite{betz2006neuronal}. {The mathematical analysis of such stochastic processes, however, is highly non-trivial. 

A particularly important quantity of interest} is the \emph{first-passage time} (FPT), that is, the {random} time it takes the process to first cross a certain threshold \cite{d2005first, condamin2007first}. FPT distributions play a crucial role both in the theory of stochastic processes and in their applications across various disciplines as they {allow} us to investigate quantitatively the uncertainty  in the emergence of system properties within a finite time horizon. For example, the time it takes cells to respond to external signals by expressing certain genes may be modelled as a FPT problem. Different characteristics of this first time distribution, for example { the variance of the FPT}, may represent evolutionarily different strategies that organisms adopt to filter fluctuations in the environment \cite{kussell2005phenotypic, tuanase2006signal, kobayashi2010implementation}. 
Examples from other disciplines include the extinction time of diseases in epidemic models, or the time {it takes to form a certain number of polymers in polymerisation processes.}

{
FPTs for stochastic processes have been of interest in statistical physics for many decades \cite{redner2001guide}.  For certain random walk or spatial diffusion processes analytic solutions have been derived \cite{benichou2011intermittent,bray2013persistence, aurzada2015persistence}. Recently, analytic results have been found for effective one-dimensional diffusion processes to a target \cite{godec2016universal, godec2016first,godec2017first}. For multi-dimensional diffusion processes to small targets approximate solutions have been derived using singular perturbation methods and matched asymptotic expansions \cite{singer2006narrow, holcman2014narrow, isaacson2016uniform, bressloff2017residence}.
}

{
The problem of computing FPT distributions for reaction processes modelled by master equations, however, is  much less explored. Generally, no tractable evolution equations exist except for one-variable, one-step processes \cite{Gardiner2009, redner2001guide}, or  certain linear and/or catalytic processes \cite{bel2009simplicity, munsky2009specificity,grima2016exact}. {For single-time properties of the underlying master equation efficient approximation methods exist relying on continuous state spaces  \cite{schnoerr2017approximation}, but it is not clear how to extend them for the computation of FPTs. Spectral methods constitute efficient approximations for small systems \cite{deuflhard2008adaptive, kazeev2014direct}. Since these methods typically scale with the size of the state space, they are not applicable to large systems.}
Some existing FPT approaches for master equations consider rare events in single-species systems and/or mean FPTs only \cite{assaf2008population,assaf2010extinction,be2016rare,weber2017master}.}

In this article, we approach the problem of computing FPTs from a novel perspective. We show that the FPT problem can be formulated exactly as a Bayesian inference problem. We achieve this by  introducing an auxiliary observation process that determines whether the process has crossed the threshold up to a given time.  This novel formulation allows us to derive an efficient approximation scheme that relies on the solution of a small set of ordinary differential equations. We will use this approximation to analyse the FPT distributions in several non-trivial examples. {We focus on reaction networks with discrete state spaces modelled  by master equations, but the developed method can also be applied to processes with continuous state spaces modelled by Fokker-Planck equations}.

The standard approach to compute the FPT of a process $\bm{x}_t$ to leave a certain region $C$ is to compute the \emph{survival probability} $Z_{[0,t]}$, that is, the probability that the process remains in $C$ on the time interval $[0,t]$ \cite{redner2001guide}. The FPT distribution is then given by the negative time derivative of $Z_{[0,t]}$. The latter can be written as a path integral  over the process with absorbing boundary of $C$ \cite{redner2001guide}. {Equivalently, one can reweigh the unconstrained process by an indicator function $p(C_{[0,t]} | \bm{x}_{[0,t]}) $} on the paths $\bm{x}_{[0,t]}$ such that $p(C_{[0,t]} | \bm{x}_{[0,t]}) = 1$ if ${\bm{x}_{\tau}} \in C$ for ${\tau} \in [0,t]$ and zero otherwise.
{One} can then write the survival probability $Z_{[0,t]}$ up till time $t$ as a path integral over the density $p(\bm{x}_{[0,t]})$  of the unconstrained process as 
\begin{align}\label{marginal_lh_general}
  Z_{[0,t]}
  & =
    \int \mathcal{D} \bm{x}_{[0,t]} \: p(\bm{x}_{[0,t]}) p(C_{[0,t]} | \bm{x}_{[0,t]}).
\end{align}
{ At the heart of our method lies the interpretation of $p(C_{[0,t]} | \bm{x}_{[0,t]})$ as a \emph{binary observation process}: an observer external to the system assesses if the process has left the region of interest or not. In this interpretation, the survival probability $Z_{[0,t]}$ constitutes the \emph{marginal likelihood} of this auxiliary observation process. The problem of computing $Z_{[0,t]}$ and hence the FPT distribution is thus formally equivalent to a \emph{Bayesian inference problem}. Note, however, that there are no experimental data involved and no data are being simulated.}

{ Moreover, note that so far no approximations have been made and \eqref{marginal_lh_general} is exact.} However, it is not obvious how to compute the path integral in \eqref{marginal_lh_general}. 
To make progress, we approximate the continuous-time process with paths $\bm{x}_{[0,t]}$ by a discrete-time version $(\bm{x}_{t_0}, \ldots,  \bm{x}_{t_N}) $ at points  $t_0=0, \ldots, t_N=t$ with spacing $\Delta t =t/N$. { The effects of such a discretisation of time on certain survival probabilities has recently been studied in \cite{majumdar2001persistence}. We will later take the continuum limit $\Delta t \to 0$ and are hence not concerned with such effects.}

This means that the \emph{global observation process} $p(C_{[0,t]} | \bm{x}_{[0,t]})$ can be written as a product  of \emph{local observation processes} {$p(C_{t_i} | \bm{x}_{t_i})$} as
\begin{align}
\label{global_constraint_as_product}
  p(C_{[0,t]} | \bm{x}_{t_0}, \ldots,  \bm{x}_{t_N}) 
  & =
    \prod_{i=0}^N p(C_{t_i} | \bm{x}_{t_i}), 
\end{align}
where $p(C_{t_i} | \bm{x}_{t_i})=1$ if $\bm{x}_{t_i} \in C$ and zero otherwise. This gives the model a Markovian structure and allows us to cast it into a sequential Bayesian inference problem, as follows.
First, we approximate the binary observation factors in \eqref{global_constraint_as_product} by a smooth approximation of the form
\begin{align}\label{soft_constraint}
  p(C_{t_i} | \bm{x}_{t_i})
  & \approx
    \exp \left(- \Delta t ~ U(\bm{x}_{t_i}, t_i) \right),
\end{align}
where $U(\bm{x}_{t_i}, t_i)$ is a smooth function that is large for $\bm{x}_{t_i} \notin C$ and close to zero for $\bm{x}_{t_i} \in C$, with a sharp slope at the boundary. { Moreover, we require $U(\bm{x}_{t_i},t_i)$ to have a tractable expectation w.r.t. a normal distribution}. The smooth approximation in \eqref{soft_constraint} proves computationally expedient in the algorithm below
 and will allow us to take the continuum limit $\Delta t \to 0$. Note that this approximation is equivalent to approximating the global binary constraint with the global soft (that is, continuous) constraint
\begin{align}
  p(C_{[0,t]} | \bm{x}_{[0,t]})
  & =
    \exp \left(- \int_0^t d \tau ~ U(\bm{x}_{\tau}, \tau) \right).
\end{align}
The survival probability  $Z_{[0,t]}$ in \eqref{marginal_lh_general} for the discrete-time system factorises as 
\begin{align}\label{marginal_lh_factorisation}
  Z_{[0,t]} \approx p(C_{t_0}) \prod_{i=0}^{N-1} p(C_{t_{i+1}} \vert C_{ \leq t_i}),
\end{align}
where  {$p(C_{t_0})$  is the probability of being in $C$ at time $t_0$} { and $p(C_{t_{i+1}} \vert C_{ \leq t_i}) \equiv p(C_{t_{i+1}} \vert C_{t_i}, C_{t_{i-1}},\ldots, C_{t_0}) = \int d\bm{x}_{t_{i+1}} p(C_{t_{i+1}} | \bm{x}_{t_{i+1}}) p(\bm{x}_{t_{i+1}} | C_{\leq t_{i}})$ is the probability that the process is found to be in $C$ at time $t_{i+1}$, given that it was in $C$ for all previous time points.}
The computation of these factors corresponds to a sequential Bayesian inference problem which can be solved { by iteratively (i) solving the master equation forward between measurement points and (ii) updating the distribution using the observation model.  More specifically, the two steps comprise }
\begin{enumerate}[(i)]
 \item Suppose we know $p(\bm{x}_{t_i} | C_{ \leq t_i}) \equiv p(\bm{x}_{t_i} |C_{t_i}, C_{t_{i-1}},\ldots, C_{t_0})$ { at time $t_i$, that is, the marginal distribution of the process at time $t_i$  conditioned on the current and all previous observations. Suppose further that using this as the initial distribution, we can solve the system (the master equation) forward in time until time point $t_{i+1}$ to obtain $p(\bm{x}_{t_{i+1}} | C_{ \leq t_i})$, that is, the marginal distribution of the process at time $t_{i+1}$  conditioned on previous observations (note that $p(\bm{x}_{t_{i+1}} | C_{ \leq t_i})$ does not include the observation $C_{t_{i+1}}$ at time $t_{i+1}$).  }
 
 \item {To obtain $p(\bm{x}_{t_{i+1}} | C_{ \leq t_{i+1}})$ we need to take the observation $p(C_{t_{i+1}} | \bm{x}_{t_{i+1}})$ at time point $t_{i+1}$ into account. This is achieved by means of Bayes' theorem as
\small
\begin{align}\label{bayes_update}
  p(\bm{x}_{t_{i+1}} | C_{\leq t_{i+1}})
  & =
    \frac{p(C_{t_{i+1}} | \bm{x}_{t_{i+1}}) p(\bm{x}_{t_{i+1}} | C_{\leq t_{i}})}{Z_{t_{i+1}}},
\end{align}
\normalsize
where we defined the normalisation $Z_{t_{i+1}}=p(C_{t_{i+1}} \vert C_{\leq t_i})$. Note that the latter is just a factor in \eqref{marginal_lh_factorisation}. }
\end{enumerate}
Performing steps (i) and (ii) iteratively from $t_0$ to $t_N$ {and keeping track of the normalisation factors in \eqref{bayes_update}} one can thus, in principle, compute the survival probability according to \eqref{marginal_lh_factorisation}.

However, steps (i) and (ii) are generally intractable, and we propose an approximation method in the following. For step (i), we need to solve the system forward in time. We do this approximately by means of the \emph{normal moment closure} \cite{Goodman1953, schnoerr2014validity, schnoerr2015comparison}, which { approximates the discrete process by a continuous one and} assumes the single-time probability distribution to be a multivariate normal distribution $\mathcal{N}(\bm{x}_t ; \bm{\mu}_t, \bm{\Sigma}_t)$ with mean  $\bm{\mu}_t$ and covariance $\bm{\Sigma}_t$. { Using this assumption on the master equation} leads to { a closed set of ordinary differential equations} for $\bm{\mu}_t$ and $\bm{\Sigma}_t$ which can be solved numerically \cite{schnoerr2017approximation}.

Now suppose that we have solved the system forward from time $t$ to $t+\Delta t$ using normal moment closure to obtain $\hat{\bm{\mu}}_{t+\Delta t}$ and $\hat{\bm{\Sigma}}_{t+\Delta t}$ and hence the distribution $p(\bm{x}_{t+\Delta t} | C_{\leq t}) = \mathcal{N}(\bm{x}_{t+\Delta t} ; \hat{\bm{\mu}}_{t+\Delta t}, \hat{\bm{\Sigma}}_{t+\Delta t})$ (step (i)). { We next have to perform the observation update in \eqref{bayes_update} in step (ii) to obtain $p(\bm{x}_{t+\Delta t} | C_{\leq t+\Delta t})$. In order to be able to use normal moment closure again in the next (i) step, we approximate $p(\bm{x}_{t+\Delta t} | C_{\leq t+\Delta t})$ by a multivariate normal distribution with mean $\bm{\mu}_{t+\Delta t}$ and covariance $\bm{\Sigma}_{t+\Delta t}$ of the r.h.s.~in \eqref{bayes_update}. } This approach is known as \emph{assumed-density filtering} in the statistics literature \cite{maybeck1982stochastic}. { In summary, with the described approximations, steps (i) and (ii) comprise
\begin{enumerate}[(i)]
\item Solve normal moment closure equations for $\bm{\mu}_t$ and $\bm{\Sigma}_t$ from time $t$ to $t + \Delta t$ to obtain $\hat{\bm{\mu}}_{t + \Delta t}$ and $\hat{\bm{\Sigma}}_{t + \Delta t}$, where $\bm{\mu}_t$ and $\bm{\Sigma}_t$ are respectively the mean and covariance of the approximating normal distribution $\mathcal{N}(\bm{x}_t ; \bm{\mu}_t, \bm{\Sigma}_t)$.
 
 \item Compute the mean $\bm{\mu}_{t + \Delta t}$ and covariance $\bm{\Sigma}_{t + \Delta t}$  of the r.h.s.~of \eqref{bayes_update} and approximate $p(\bm{x}_{t + \Delta t} | C_{\leq t + \Delta t})$ in \eqref{bayes_update} with a corresponding normal distribution $\mathcal{N}(\bm{x}_{t + \Delta t} ; \bm{\mu}_{t + \Delta t}, \bm{\Sigma}_{t + \Delta t})$.
 \end{enumerate}
}

{ Next, we derive a continuous time description combining steps (i) and (ii)}. { This is achieved by first expanding the update in step (i) leading from $\bm{\mu}_t$ and $\bm{\Sigma}_t$ to $\hat{\bm{\mu}}_{t + \Delta t}$ and $\hat{\bm{\Sigma}}_{t+\Delta t}$ in $\Delta t$, which gives a single Euler step update of the moment closure equations. Similarly, we expand step (ii) which leads from $\hat{\bm{\mu}}_{t + \Delta t}$ and $\hat{\bm{\Sigma}}_{t+\Delta t}$ to $\bm{\mu}_{t + \Delta t}$ and $\bm{\Sigma}_{t+\Delta t}$, as follows. Note first that, by definition, the normalisation $Z_{t+\Delta t}$ in \eqref{bayes_update} can be written as
\small
\begin{equation}\begin{split}\label{marginal_lh_discrete}
   Z_{t+\Delta t}
  & \approx
    \int d\bm{x} \mathcal{N}(\bm{x}; \hat{\bm{\mu}}_{t+\Delta t}, \hat{\bm{\Sigma}}_{t+\Delta t}) 
     e^{-  \Delta t U(\bm{x}, t+\Delta t)}.
\end{split}\end{equation}
\normalsize
Taking the logarithm of both sides, expanding in $\Delta t$ and taking derivatives w.r.t. $\hat{\bm{\mu}}_t$ and $\hat{\bm{\Sigma}}_t$, one can derive the desired expansion of the update in step (ii). The resulting expansions of steps (i) and (ii) can then be combined to give unifying update equations for $\bm{\mu}_t$ and $\bm{\Sigma}_t$ (see Supplemental Material \cite{supp} for a derivation). 
Finally, taking the continuum limit $\Delta t \to 0$ gives the following closed set of differential equations}
\small
\begin{align}\label{adf_equation_mean}
  \frac{\partial}{\partial t}
  \bm{\mu}_t
   & =
    \left(\frac{\partial}{\partial t} 
    \bm{\mu}_t\right)^{\text{MC}}    
    - \bm{\Sigma}_t
        \frac{\partial}{\partial \bm{\mu}_t} \langle U(\bm{x}_t, t) \rangle_{\mathcal{N}(\bm{x}_t ; \bm{\mu}_t, \bm{\Sigma}_t)},\\
\label{adf_equation_covariance}
  \frac{\partial}{\partial t}
  \bm{\Sigma}_t
   & =
    \left( \frac{\partial}{\partial t} 
      \bm{\Sigma}_t \right)^{\text{MC}} 
    - 2 \bm{\Sigma}_t \left( \frac{\partial}{\partial\bm{\Sigma}_t} \langle U(\bm{x}_t, t) \rangle_{\mathcal{N}(\bm{x}_t ; \bm{\mu}_t, \bm{\Sigma}_t)} \right) \bm{\Sigma}_t, \\
\label{adf_equation_partition_function}
  \frac{\partial}{\partial t} & \log Z_{[0,t]}
   =
  -  \langle  U(\bm{x}_{t}, t) \rangle _{\mathcal{N}(\bm{x}_t ; \bm{\mu}_t, \bm{\Sigma}_t)}.
\end{align}
\normalsize
Here, the first terms on the r.h.s.~of \eqref{adf_equation_mean} and \eqref{adf_equation_covariance} are respectively the equations for the mean and covariance as obtained from the normal moment closure (MC) for the unconstrained system, while the second terms incorporate the auxiliary observation. Equation \eqref{adf_equation_partition_function} gives the desired survival probability for the process. We term this method for computing FPT distributions {\it Bayesian First-Passage Times} (BFPT).

Equations \eqref{adf_equation_mean}-\eqref{adf_equation_partition_function} are the central result of this article. They constitute closed form ordinary differential equations for the mean, covariance and log-survival probability of the process, for which efficient numerical integrators exist. Solving these equations forward in time on an interval $[0,t]$ provides an approximation of the survival probability {$ Z_{[0,\tau]}$} for all ${\tau} \in [0,t]$ (on the time grid of the numerical integrator), from which the FPT distribution $p(\tau; C)$ can be derived for all $\tau \in [0,t]$ by taking the negative derivative of {$Z_{[0,\tau]}$, that is, $p(\tau; C) = - \partial Z_{[0,\tau]}/\partial \tau $. The number of equations scales with the square of the number of species, and the method hence is applicable to large systems. }{Crucially, and in contrast to stochastic simulations and spectral methods, the complexity of the method is independent of the population size and the size of the state space.}
Similar equations were obtained in a different context in \cite{cseke2013approximate, cseke2016expectation}  by means of a variational approximation. 

In the derivation of \eqref{adf_equation_mean}-\eqref{adf_equation_partition_function} we utilised three approximations:~{after discretising time,} we approximated the unconstrained process using normal moment closure and the observation updates by projections onto a normal distribution. {We then approximated the binary observation model by a soft loss function,  which allowed us to take the continuum limit in time}. Depending on the problem, the relative contribution of the three sources to the overall error may vary.

\begin{figure}[t]
\includegraphics[width=0.49\textwidth]{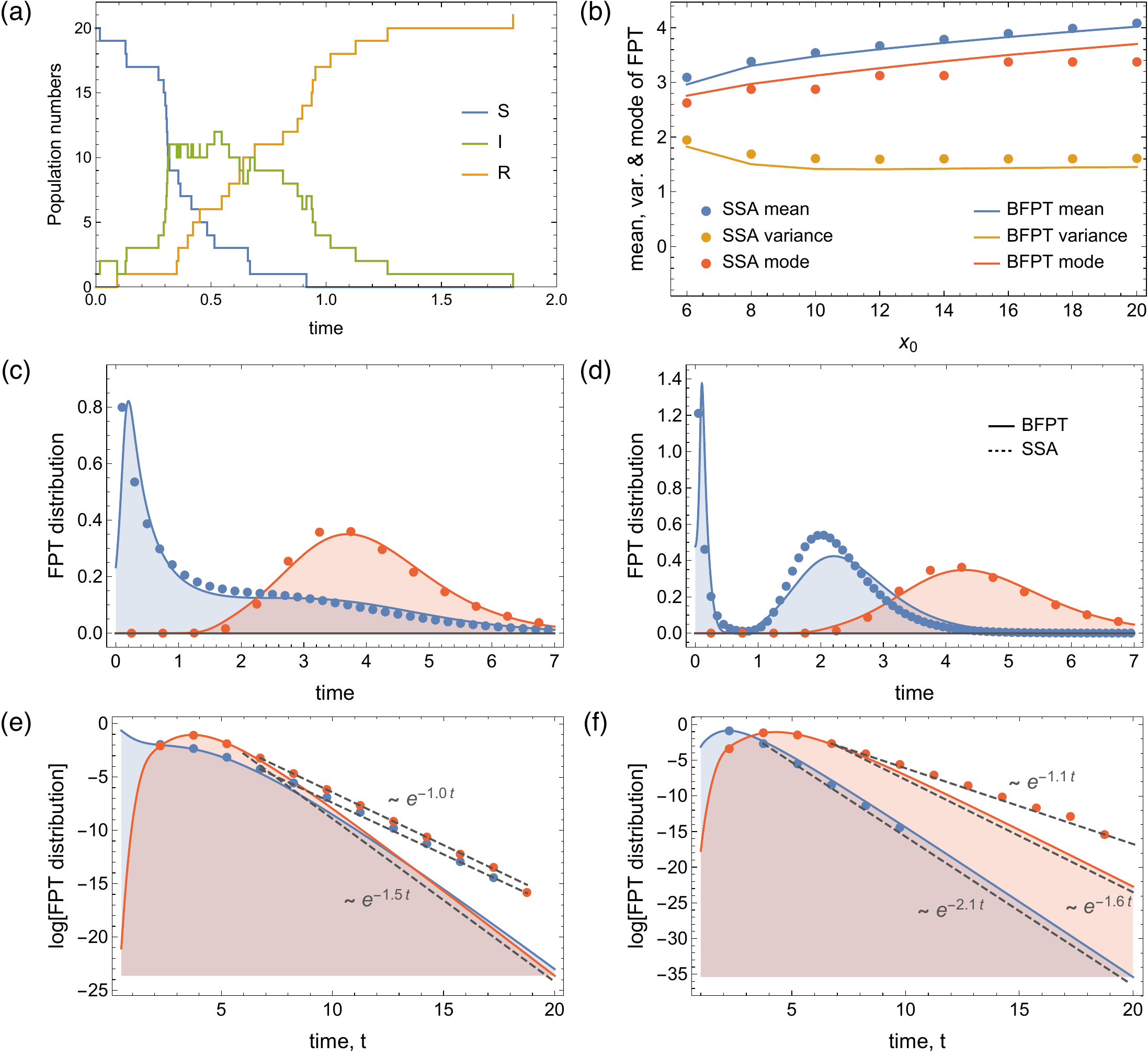}
\caption{Results for the epidemic system \eqref{sir_interactions}. (a) Simulated path of the process. (b) Mean, variance and  mode of the FPT distribution of species $I$ becoming extinct as a function of the initial populations $x_0$ of species $S$, from the stochastic simulation algorithm (SSA, dots, $10^6$ samples per point) and from BFPT (lines). The rate constants are set to $k_1 = 0.5$ and $k_2 = 1$, and the initial value for species $I$ is set to $y_0 = 2 x_0$.
(c),(d) FPT distributions as obtained from the SSA (dots, $10^7$ samples per parameter set) and BFPT (lines) for the parameter set $(x_0, y_0, k_1, k_2)$ chosen as $(6,1, 0.25,1)$ (blue, (c)),  $(20, 10 , 0.5, 1)$ (red, (c)), $(20, 1 , 0.5, 2 )$ (blue, (d)), and $(40, 10, 0.25, 1)$ (red, (d)). The parameter $a$ in \eqref{exp_potential} was chosen as $a=-3$ for the blue curve in (c) and $a=-1.5$ for all other figures. (e),(f) same results as (c),(d) but logarithmic scale.}
\label{sir_pic}
\end{figure}

{The choice of loss function  $U(\bm{x}, t)$ depends on the problem at hand. In general, for computational convenience one needs to be able to compute analytically the expectation of the loss function w.r.t.~a multivariate normal distribution. In our examples, we use an exponential loss function to constrain the $i^{\text{th}}$ component of the state vector $\bm{x}$ about a threshold c
\begin{equation}\begin{split}\label{exp_potential}
  U(\bm{x}, t)
  & =
     \exp (a(x^i-c)), \quad a \in \mathbb{R}, c \in \mathbb{R}. 
\end{split}\end{equation}
The absolute value of the parameter $a$ determines the steepness of the loss function. In principle, we choose $a$ as large as numerically feasible. For a detailed discussion on the choice of loss function see Supplemental Material \cite{supp}.
}

We now examine the performance of BFPT on three examples. { For the analytically tractable Poisson birth process we find that BFPT captures the low-order moments and the bulk mass of the distribution accurately while giving the correct scaling law for the tail of the distribution (see Supplemental Information for details).}

Next, we consider an epidemic system consisting of a susceptible population $S$, an infected population $I$ and a recovered population $R$ and interactions
\small
\begin{equation}\begin{split}\label{sir_interactions}
   S + I \xrightarrow {\quad k_1 \quad } 2 I,    \quad I \xrightarrow{\quad k_2 \quad} R.
\end{split}\end{equation}
\normalsize
This system is frequently modelled as a continuous-time Markov-jump process  to model a disease spreading through a population.
 $k_1$ and $k_2$ in \eqref{sir_interactions} are the corresponding rate constants. Let $\bm{x}_t=(x_t,y_t,z_t)$, where $x_t, y_t$ and $z_t$ denote the populations of $S, I$ and $R$, respectively.
We are interested in the probability distribution of time for the disease to be permanently eradicated, that is, the time it takes the process to reach a state with $y_t=0$. 

Figs.\;1(b) shows the mean, variance and mode of the FPT to extinction as obtained from our method and the stochastic simulation algorithm \cite{Gillespie1976}. We find that BFPT accurately captures the mean, variance and mode of the FPT over a wide range of varying initial values for  $S$ and~$I$.

Figs.\;1(c) and (d) show the FPT distributions for four different parameter sets. The modality, mode and overall shape of the FPT are well captured, even for highly skewed  and  bimodal distributions (c.f.~blue curve in Figs.\;\ref{sir_pic}(c) and (d), respectively). 
In some cases the method predicts less peaked distributions than actual (not shown here). {Figs.\;1(e) and (f) show the same results on logarithmic scale. We observe that our method correctly predicts an exponential scaling (straight lines in logarithmic scale), although the scaling is not always accurate, indicating a worse approximation in the tails of the distribution. }

The value of the approach is borne out by considering its computational efficiency: for the results shown in Fig.\;\ref{sir_pic}, BFPT is several orders of magnitude faster than stochastic simulations. For example, simulating $10^7$ paths to obtain the results shown in Figs.\;\ref{sir_pic}(c)-(f) takes about $10^3$-$10^4$ seconds in our  implementation of the direct stochastic simulation algorithm \cite{Gillespie1976}, whereas BFPT takes less than a second.

\begin{figure}[t]
\includegraphics[width=0.5\textwidth]{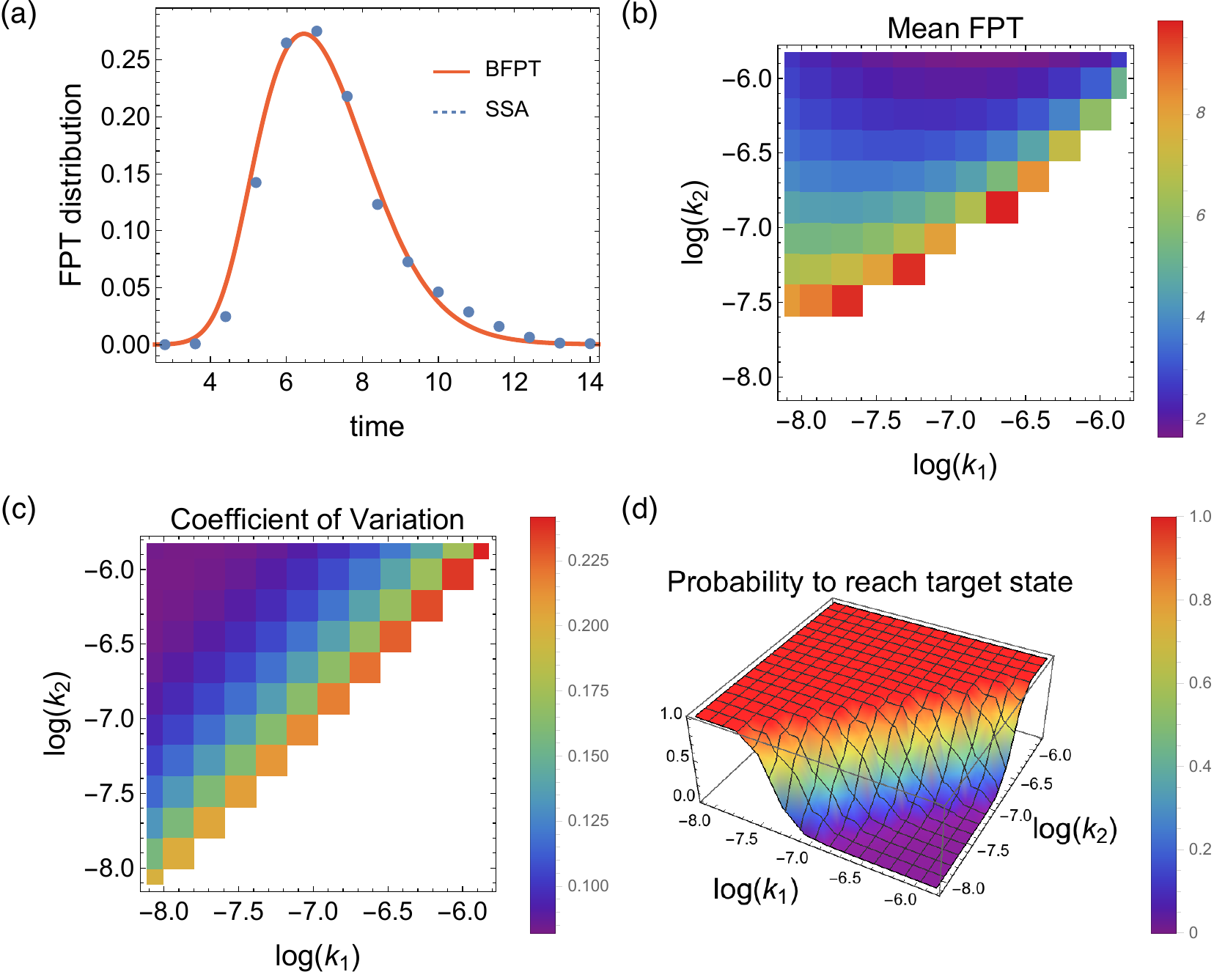}
\caption{Results for the polymerisation system in \eqref{poly_interactions}. (a) FPT distribution for the parameters $k_1 = k_2 = 10^{-3}$ obtained from the stochastic simulation algorithm (SSA, dots, $10^4$ samples).
 (b),(c) Heat plots of the mean and the coefficient of variation (defined as standard deviation divided by mean) of the FPT to produce $200$ trimers starting with $10^3$ monomers, as a function of $k_1$ and $k_2$ on logarithmic scale. (d) Corresponding 3D plot for the normalisation of the FPT distribution, that is, the probability with which at least $200$ trimers are being produced. The white areas in (b) and (c) indicate that either the value is larger than the plotted range or that the target state is reached with such small probability that an estimation of moments is not sensible. The parameter $a$ in \eqref{exp_potential} was fixed to $a=0.2$ for all figures. }
\label{fig_trimerisation}
\end{figure}

Finally, we apply BFPT to a polymerisation system of monomers $X$, dimers $XX$ and trimers $XXX$ with interactions 
\small
\begin{equation}\begin{split}\label{poly_interactions}
   X + X \xrightarrow { k_1  } XX,    \quad XX + X  \xrightarrow{ k_2 } XXX.
\end{split}\end{equation}
\normalsize
Starting from a fixed number of $10^3$ of monomers, zero dimers and zero trimers, we are interested in the FPT it takes to produce $200$ trimers. We are interested in exploring the dependence of this FPT distribution on the parameters of the system (dimerisation and trimerisation rate $k_1$ and $k_2$, respectively); such parameter exploration is computationally too demanding to be performed by brute force simulation without access to dedicated hardware since the FPT distribution needs to be estimated for a large number of parameter sets.

Fig.~\ref{fig_trimerisation} shows the results for this process. We observe excellent agreement between BFPT and simulations for a particular value of the parameters (Fig.~\ref{fig_trimerisation}~(a)). The 
heat plot for the mean as a function of $k_1$ and $k_2$ indicates that for a given trimerisation rate $k_2$ a minimal mean FPT is achieved for an intermediate value of dimerisation rate $k_1$ (Fig.~\ref{fig_trimerisation}~(b)). We find a linear relationship $k_2 \approx 2.3 k_1$ between the two rates for the location of these minima. The variance of FPT behaves quantitatively similarly (not shown in the figure). The coefficient of variation (Fig.~\ref{fig_trimerisation}~(c)), however, becomes minimal for small values of $k_1$ for a given $k_2$. This reveals an unexpected tradeoff between an optimal mean FPT and optimal noise-to-mean ratio (coefficient of variation). Fig.~\ref{fig_trimerisation}~(d) shows the probability that the target state is reached, that is, the probability that at least $200$ trimers are being produced. We find that there are two parameter regions, one with probability close to one and one with probability close to zero, and a small transition range between these two with boundary $k_2 \approx 0.55 k_1$. 

In conclusion, we have shown that the problem of computing survival probabilities and FPT distributions for stochastic processes can be formulated as a sequential Bayesian inference problem. This novel formulation opens the way for a new class of efficient approximation methods from machine learning and computational statistics to address this classical intractable problem. Here, we derived an approximation { for FPT distributions which relies on solving a small set of ordinary differential equations. This results in considerable efficiency gains; empirically, we found the approximation to be highly accurate in several examples. However, we do not have at present systematic error estimates for the method; we leave the investigation of such bounds and possible correction methods for future work. In particular, it will be interesting to study the tail behaviour of FPT distributions with our method, as these were not always captured well in our examples. We notice that, while we applied our method to processes with discrete state spaces modelled by master equations, in principle, it can equally easily be applied to processes with continuous state spaces modelled by Fokker-Planck equations.}

This work was supported by the Leverhulme Trust [RPG-2013-171]; and the European Research Council [MLCS 306999]. We thank Manfred Opper for insightful discussions.

\bibliography{fpt}

\appendix
\small
%%%%%%%%%%%%%%%%%%%%%%%%%%%%%%
%%%%%%%%%%%%%%%%%%%%%%%%%%%%%%
%%%%%%%%%%%%%%%%%%%%%%%%%%%%%%
\section{Derivation of main result}

Here, we derive the main results of this work given in Equations (8)-(10). To this end, consider step (i) of the sequential scheme presented in the main text and suppose we have solved the corresponding moment closure equations from $t$ to $t+\Delta t$ to obtain $\hat{\bm{\mu}}_{t+\Delta t}$ and $\hat{\bm{\Sigma}}_{t+\Delta t}$ which we can thus expand as
\begin{align}\label{mc_update_supp1}
  \hat{\bm{\mu}}_{t+\Delta t}
  & =
    \bm{\mu}_t +  \Delta t \left(\frac{\partial}{\partial t}  \bm{\mu}_t\right)^{\text{MC}}   + O(\Delta t^2), \\
\label{mc_update_supp2}
  \hat{\bm{\Sigma}}_{t+\Delta t}
  & =
    \bm{\Sigma}_t +\Delta t  \left(\frac{\partial}{\partial t}  \bm{\Sigma}_t\right)^{\text{MC}}   + O(\Delta t^2).
\end{align}
$\hat{\bm{\mu}}_{t+\Delta t}$ and $\hat{\bm{\Sigma}}_{t+\Delta t}$ are the moments of the distribution at time $t+\Delta t$  prior to the update in (6):
\begin{align}\label{prior_supp}
  p(\bm{x}_{t+\Delta t} | C_{\leq t}) = \mathcal{N}(\bm{x} ; \hat{\bm{\mu}}_{t+\Delta t}, \hat{\bm{\Sigma}}_{t+\Delta t}).
\end{align}
Note that $p(\bm{x}_{t+\Delta t} | C_{\leq t})$ does not include the observation $C_{t + \Delta t}$.
The latter can be taken into account using update equation (6). Using the exponential form of the constraint given in (3) one finds that the normalisation in (6) can be written as stated in (7). Using this, one can derive the relations
\begin{align}\label{logZ_derivative_1}
  Z_{t + \Delta t}
  & =
    \int d \bm{x}_{t+\Delta t}  e^ {- \Delta t ~ U(\bm{x}_{t+\Delta t}, t+\Delta t)}\\
  & \nonumber \quad \quad \quad
    \times \mathcal{N}(\bm{x}_{t+\Delta t}; \hat{\bm{\mu}}_{t+\Delta t}, \hat{\bm{\Sigma}}_{t+\Delta t}),  \\
\label{logZ_derivative_2}
   \frac{\partial}{\partial \hat{\bm{\mu}}}\log Z_{t+\Delta t}  
  & \\ = 
  \nonumber 
     \langle\frac{\partial}{\partial \hat{\bm{\mu}}} 
    \log &  \mathcal{N} (\bm{x}_{t + \Delta t}; \hat{\bm{\mu}}_{t+\Delta t}, \hat{\bm{\Sigma}}_{t+\Delta t})   
    \rangle_{p(\bm{x}_{t + \Delta t} | C_{ \leq t + \Delta t})}, \\
\label{logZ_derivative_3}
   \frac{\partial}{\partial\hat{\bm{\Sigma}}}\log Z_{t+\Delta t} 
  & \\ =
  \nonumber 
     \langle\frac{\partial}{\partial \hat{\bm{\Sigma}}}
    \log & \mathcal{N} (\bm{x}_{t + \Delta t}; \hat{\bm{\mu}}_{t+\Delta t}, \hat{\bm{\Sigma}}_{t+\Delta t})   
    \rangle_{p(\bm{x}_{t + \Delta t} | C_{ \leq t + \Delta t})}, 
\end{align}
where we have used 
\small
\begin{equation}\begin{split}
  p(&\bm{x}_{t + \Delta t}  | C_{ \leq t + \Delta t})
    \\
  &   =
    \frac{e^ {- \Delta t ~ U(\bm{x}_{t+\Delta t}, t+\Delta t)} 
      \mathcal{N}(\bm{x}_{t + \Delta t}; \hat{\bm{\mu}}_{t+\Delta t}, \hat{\bm{\Sigma}}_{t+\Delta t}) }
    {Z_{t + \Delta t}},
\end{split}\end{equation}
which follows from (3) and (6).
Taking the logarithm of \eqref{logZ_derivative_1} and expanding in $\Delta t$ we further find 
\begin{equation}\begin{split}\label{marginal_lh_discrete_expansion_supp}
  & \log  Z_{t+\Delta t}
 \\
    & \quad =
    - \Delta t \langle U(\bm{x}, t+\Delta t) \rangle_{\mathcal{N}(\bm{x} ; \hat{\bm{\mu}}_{t+\Delta t}, \hat{\bm{\Sigma}}_{t+\Delta t})} + O(\Delta t^2).
\end{split}\end{equation}
Using  \eqref{marginal_lh_discrete_expansion_supp} in the l.h.s. of \eqref{logZ_derivative_2} and \eqref{logZ_derivative_3} and performing the logarithm and derivative on the r.h.s., we obtain the update equations 
\begin{align}\label{marginal_lh_discrete_derivatives_supp}
  & \bm{\mu}_{t+\Delta t}
   =
    \hat{\bm{\mu}}_{t+\Delta t} \\
  \nonumber & \quad 
   - \Delta t  \hat{\bm{\Sigma}}_{t+\Delta t} \frac{\partial}{\partial \hat{\bm{\mu}}} \langle U(\bm{x}, t+\Delta t) \rangle_{\mathcal{N}(\bm{x} ; \hat{\bm{\mu}}_{t+\Delta t}, \hat{\bm{\Sigma}}_{t+\Delta t})} + O(\Delta t^2), \\
\label{marginal_lh_discrete_derivatives2_supp}
  & \bm{\Sigma}_{t+\Delta t}
   = 
    \hat{\bm{\Sigma}}_{t+\Delta t}\\
  \nonumber & \quad
    - 2 \Delta t  \hat{\bm{\Sigma}}_{t+\Delta t}  \frac{\partial \langle U(\bm{x}, t+\Delta t) \rangle_{\mathcal{N}(\bm{x} ; \hat{\bm{\mu}}_{t+\Delta t}, \hat{\bm{\Sigma}}_{t+\Delta t})} }{\partial \hat{\bm{\Sigma}}_{t+\Delta t}}  \hat{ \bm{\Sigma}}_{t+\Delta t} \\
    \nonumber & \quad \quad
      + O(\Delta t^2).
\end{align}
Plugging \eqref{mc_update_supp1} or \eqref{mc_update_supp2} into \eqref{marginal_lh_discrete_derivatives_supp} or \eqref{marginal_lh_discrete_derivatives2_supp} and expanding in $\Delta t$ we find
\begin{align}
   \bm{\mu}_{t+\Delta t}
   & =
    \bm{\mu}_t + \Delta t  \left(\frac{\partial}{\partial t}  \bm{\mu}_t\right)^{\text{MC}} \\
   & \nonumber \quad \quad
   - \Delta t ~\bm{\Sigma}_{t} \frac{\partial}{\partial \bm{\mu}} \langle U(\bm{x},  t) \rangle_{\mathcal{N}(\bm{x} ; \bm{\mu}_{t}, \bm{\Sigma}_{t})} + O(\Delta t^2), \\
   \bm{\Sigma}_{t+\Delta t}
   & =
    \bm{\Sigma}_t + \Delta t  \left(\frac{\partial}{\partial t}  \bm{\Sigma}_t\right)^{\text{MC}} \\
    & \nonumber \quad \quad
    - 2 \Delta t~\bm{\Sigma}_{t} \left( \frac{\partial}{\partial \bm{\Sigma}_{t}}  \langle U(\bm{x}, t) \rangle_{\mathcal{N}(\bm{x} ; \bm{\mu}_{ t}, \bm{\Sigma}_{t})} \right)\bm{\Sigma}_{t} + O(\Delta t^2).
\end{align}
Taking the continuum limit $\Delta t \to 0$ we obtain (8) and (9). Similarly, using  that we can write 
\begin{align}\label{eq2}
   \log  Z_{t+\Delta t} = \log  Z_{[0,t+\Delta t]} -  \log Z_{[0,t]},
\end{align}
in \eqref{marginal_lh_discrete_expansion_supp} we obtain
\begin{equation}\begin{split}
  & \log  Z_{[0,t+\Delta t]} -  \log Z_{[0,t]} \\
  &  \quad
=
    - \Delta t \langle U(\bm{x}, t+\Delta t) \rangle_{\mathcal{N}(\bm{x} ; \hat{\bm{\mu}}_{t+\Delta t}, \hat{\bm{\Sigma}}_{t+\Delta t})} + O(\Delta t^2),
\end{split}\end{equation}
which is just the discrete-time version of (10).

%%%%%%%%%%%%%%%%%%%%%%%%%%%%%%
%%%%%%%%%%%%%%%%%%%%%%%%%%%%%%
%%%%%%%%%%%%%%%%%%%%%%%%%%%%%%
\section{Choice of  loss function}\label{app_sim_details}

We approximate the binary observation process by a soft constraint of exponential form (3). For the studied examples, we choose an exponential loss function (11). This (softly) confines the process to the interval $x_t \in [ - \infty , c] ([c, \infty])$ if $a$ is positive (negative). 
This leaves us with the choice of the magnitude of $a$. A larger $|a|$ corresponds to a steeper loss function and hence to a better approximation of the binary observation process. However, a larger $|a|$ also leads to stronger non-linearity of the equations (8)-(10). This will typically limit $|a|$ for numerical feasibility. Optimally, one would hope the results to converge in $|a|$ for values for which the equations are still numerically feasible. For some of the examples studied we indeed found this to be the case. For others, we chose $|a|$ as large as numerically feasible. A more systematic way of choosing $|a|$ is left for future work.

%%%%%%%%%%%%%%%%%%%%%%%%%%%%%%
%%%%%%%%%%%%%%%%%%%%%%%%%%%%%%
%%%%%%%%%%%%%%%%%%%%%%%%%%%%%%
\section{Poisson process}\label{app_birth_process}

The first example in the main text is the Poisson birth process comprising a single species $X$ and a single reaction 
\begin{equation}\begin{split}
   \varnothing \xrightarrow {\quad k \quad } X.
\end{split}\end{equation}
If there are zero $X$ molecules in the system initially it is easy to show that the solution $p(x, t)$ of the corresponding master equation at time $t$ is given by a Poisson distribution with mean $k t$:
\begin{equation}\begin{split}
    p(x, t) = \frac{e^{-kt} (kt)^x}{x!},
\end{split}\end{equation}
where $x \in \mathbb{N}$ is the number of $X$ molecules in the system. Suppose we want to compute the FPT distribution $p(\tau ; c)$ for reaching a state $c \in \mathbb{N}_+$. Since the number of $X$ molecules never decreases, $p(\tau ; c)$ is simply given by the probability of being in state $c-1$ times the rate $k$ of the reaction firing which means jumping into state $c$: 
\begin{equation}\begin{split}
    p(\tau; c) = k \times p(c-1, \tau) = k \times \frac{e^{-k\tau} (k\tau)^{c-1}}{(c-1)!}.
\end{split}\end{equation}
The long-time behaviour is hence given by 
\begin{equation}\begin{split}\label{poisson_process_asymptotic_fpt}
    p(\tau; c) \sim e^{-k \tau}, \quad \tau \to \infty.
\end{split}\end{equation}
Next, we derive the same result using our method. We use an exponential loss function of the form given in $(11)$ and $a>0$. Plugging the expectation of the loss function into $(8)$ and $(9)$ we obtain the evolution equations for the mean $\mu_t$ and variance $\Sigma_t$ of the process
\begin{equation}\begin{split}
  \frac{\partial}{\partial t} \mu_t 
  & =
    k - a \Sigma e^{a(\mu_t + a \Sigma / 2 -c)}, \\
  \frac{\partial}{\partial t} \Sigma_t 
  & =
    k - a^2 \Sigma^2 e^{a(\mu_t + a \Sigma / 2 -c)}.
\end{split}\end{equation}
In steady state $\frac{\partial}{\partial t} \mu_t  = \frac{\partial}{\partial t} \Sigma_t = 0$ this is solved by 
\begin{equation}\begin{split}
   \mu^*
  & =
    c - \frac{1}{2} + \frac{1}{a}\log(k), \\
   \Sigma^*
  & =
    1/a.
\end{split}\end{equation}
Plugging these into the equation (10) for the survival probablity we find 
\begin{equation}\begin{split}
  \frac{\partial}{\partial t} \log Z_{[0,t]} \big|_{\mu^*, \Sigma^*}
  & =
    - e^{a(\mu^* + a \Sigma^* / 2 -c)}
  =
    -k.
\end{split}\end{equation}
Solving this we find the long-time FPT distribution 
\begin{equation}\begin{split}
  p(\tau; c) \big|_{\mu^*, \Sigma^*} 
  & = \frac{\partial}{\partial t} Z_{[0,t]} \big|_{\mu^*, \Sigma^*, t=\tau}
   \sim 
     e^{-k \tau},
\end{split}\end{equation}
which is the same as \eqref{poisson_process_asymptotic_fpt}.
We thus find that our method predicts the exact asymptotic behavior of the FPT distribution for long times. 
\end{document}